\documentclass[aps,prb,longbibliography,reprint,amsmath,amssymb,superscriptaddress]
{revtex4-2}
\usepackage{graphicx}% Include figure files
\usepackage{dcolumn}% Align table columns on decimal point
\usepackage{bm}% bold math
\usepackage{color} % For blue in-text comments and additions
\usepackage{physics}
\usepackage{hyperref}
\usepackage{subfigure}
\hypersetup{
    colorlinks = true,
    linkcolor = blue,
    citecolor = blue,
    anchorcolor = blue,
    urlcolor = blue
    }

\def\lesssim{\ \raise.3ex\hbox{$<$}\kern-0.8em\lower.7ex\hbox{$\sim$}\ }
\def\gesim{\ \raise.3ex\hbox{$>$}\kern-0.8em\lower.7ex\hbox{$\sim$}\ }

\newcommand \beq{\begin{eqnarray}}
\newcommand \eeq{\end{eqnarray}}

\usepackage{amsmath,amssymb}
\usepackage{fixmath}
\usepackage[normalem]{ulem}

\usepackage{comment}

\begin{document}

\preprint{RIKEN-iTHEMS-Report-24}

\title{Supersymmetry-like tunneling current noise as a probe of Goldstino excitation in a Bose-Fermi mixture}

\author{Tingyu Zhang}
\email{zhangty@g.ecc.u-tokyo.ac.jp}
\affiliation{Department of Physics, School of Science, The University of Tokyo, Tokyo 113-0033, Japan}
\affiliation{Interdisciplinary Theoretical and Mathematical Sciences Program (iTHEMS), RIKEN, Wako 351-0198, Japan}

\begin{abstract}
The Goldstino, which is a fermionic Nambu-Goldstone mode, has been predicted in a Bose-Fermi mixture when the supersymmetry is broken. To detect this excitation mode, we theoretically investigate the shot noise of the supersymmetry-like tunneling current in a weakly interacting ultracold Bose-Fermi mixture. The Fano factor, which is defined by the noise-to-current ratio, reflect the elementary carriers of the tunneling process. The change of the Fano factor microscopically as the density changes evinces a crossover from the quasiparticle transport to multiparticle (Goldstino) transport. The tunneling channel can also be changed by tuning the potential barrier.
\end{abstract}

\maketitle

\section{Introduction}

The Ultracold atoms have become a unique tool to study the nature of quantum many-body systems owing to their cleanness and high experimental flexibility. The well-developed experimental techniques for cold atoms enable exploration of various regimes including weakly interacting Bose-Einstein condensates~\cite{RevModPhys.74.875,RevModPhys.81.647,aveline2020observation}, the superfluid-Mott insulator transition~\cite{greiner2002quantum,PhysRevLett.104.160403} for bosonic systems, and the crossover between the Bardeen-Cooper-Schrieffer (BCS) and Bose-Einstein condensation (BEC) regimes~\cite{PhysRevLett.92.040403,PhysRevLett.92.120401,BCS-BEC}, as well as the itinerant ferromagnetic transition~\cite{science.1177112} for the case of fermions. 

The realization of mixtures of bosonic and
fermionic quantum fluids~\cite{PhysRevLett.87.080403,science.1059318,science.1255380} has opened new avenues for investigating quantum many-body systems, where the components obey different statistical rules and interact with each other~\cite{PhysRevLett.90.170403,PhysRevLett.92.050401,PhysRevLett.96.180402}. The interspecies interaction allows the sympathetic cooling~\cite{PhysRevLett.78.586,science.1059318,PhysRevA.64.011402,PhysRevLett.87.080403,PhysRevA.105.043312}, and can change the dynamics in each components~\cite{PhysRevLett.131.083003}. issues in the study of Bose-Fermi mixtures include their stability and miscibility~\cite{PhysRevLett.80.1804,PhysRevB.78.134517,PhysRevLett.96.180402,PhysRevA.83.041603,PhysRevLett.110.115303,PhysRevLett.120.243403,PhysRevA.103.063317,PhysRevLett.132.033401}, transport properties~\cite{PhysRevA.84.033627,Wang_2020,PhysRevB.110.064512}, and the possible applications in quantum simulations to achieve supersymmetry~\cite{PhysRevLett.100.090404,PhysRevA.81.011604,PhysRevA.91.063620,PhysRevA.92.063629,PhysRevA.93.033642,PhysRevD.94.045014,PhysRevA.96.063617,PhysRevResearch.3.013035}. The supersymmetry, which assigns a superpartener (a particle with the same mass but opposite statistics) to each elementary particle, appears to be broken when there is an explicit mass or chemical potential imbalance between the two components. The symmetry breaking leads to the emergence of a massless
fermionic Nambu-Goldstone mode, referred to as a Goldstino~\cite{SALAM1974465,WITTEN1981513,LEBEDEV1989669,PhysRevLett.100.090404}. The Goldstino has garnered significant research interest, as it carries a fermionic quantum number, unlike most collective modes in cold atom systems, yet shares many similarities with the more familiar Nambu-Goldstone boson.~\cite{PhysRev.122.345,goldstone1961field}. Beyond cold atomic systems, a fermionic mode linked to spontaneous supersymmetry breaking is also predicted in quark-gluon plasmas at extremely high temperatures~\cite{HIDAKA201293,PhysRevD.87.096011}. In spite of its importance for understanding the supersymmetry, the existence of Goldstino remains elusive and has yet to be observed. However, it has been found that Goldstino excitations can contribute to the tunneling transport as a multiparticle process in an ultracold Bose-Fermi mixture~\cite{PhysRevB.110.064512}, and thus could be detected through tunneling signals. One challenge, however, is to distinguish the signal of Goldstino from those of quasiparticle processes. 
\begin{figure}[t]
    \centering
    \includegraphics[width=8.6cm]{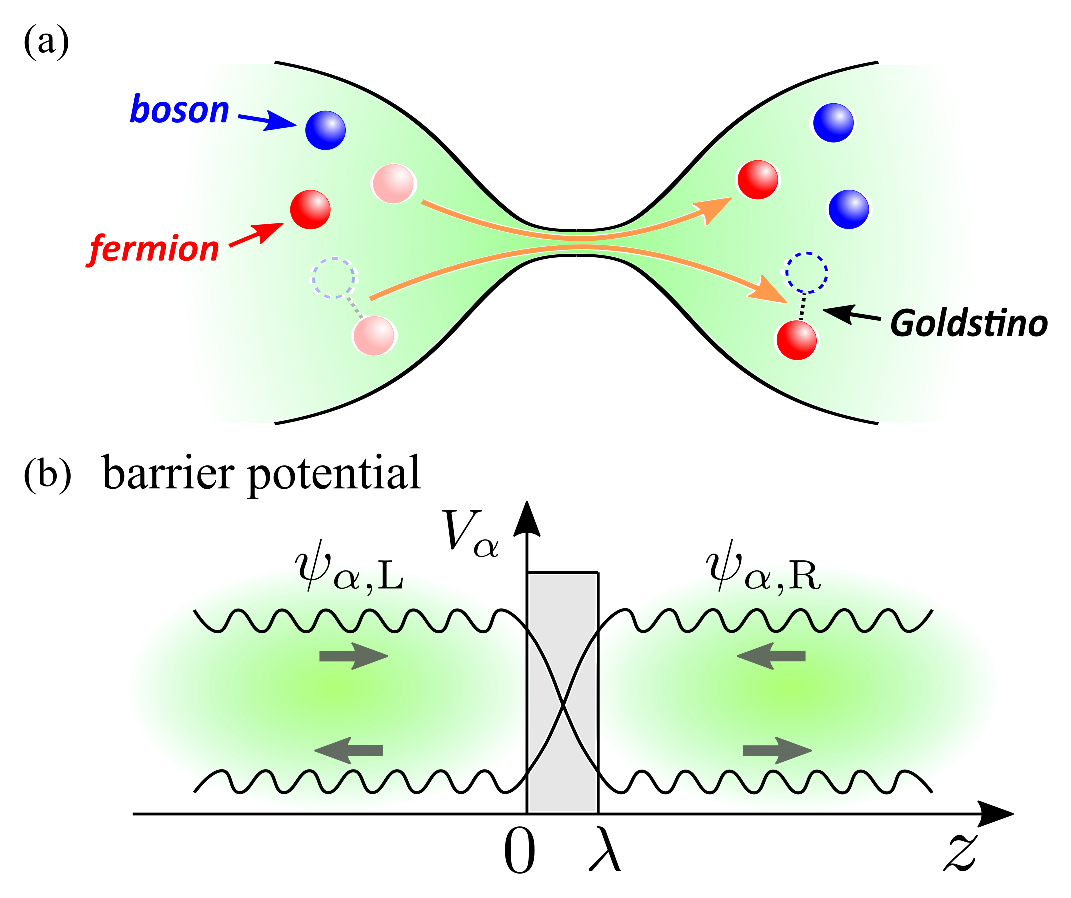}
    \caption{
    (a) Schematic view of the two-terminal system we study. Chemical potential bias between two reservoirs provokes quasiparticle as well as supercharge tunnelings. The dashed circle represents the hole, and the pair of a fermion and a bosonic hole denotes a Goldstino. (b) Rectangular potential barrier with height $V_\alpha$ ($\alpha=b$ for bosons and $\alpha=f$ for fermions) and width $\lambda$ applied for the junction. $\psi_{\alpha,{\rm L(R)}}$ denotes wave functions of particles in the left (right) reservoir.}\label{schematic}
\end{figure}

Out of consideration for this, we propose detecting the Goldstino exitation through the shot noise of the supersymmetry-like tunneling current. Quantum shot noise arises from the discrete nature of current carriers, and is therefore proportional to both the effective charge and the average current~\cite{BLANTER20001}. In solid state systems, shot noise has been used to probe the fractional charges in the fractional quantum Hall regime~\cite{PhysRevLett.79.2526,DEPICCIOTTO1998395} and the effective spin associated with magnon tunneling through a ferromagnet-insulator-normal metal interface~\cite{PhysRevLett.120.037201}. It is also shown recently that in cold atomic systems the shot noise can serve as a probe of multiparticle processes~\cite{pgad045,PhysRevApplied.21.L031001}, revealing a crossover from the quasiparticle transport to multiparticle transport as the interaction strength is tuned by using Feshbach resonances~\cite{RevModPhys.82.1225}. In our work, we focus on a small mass-balanced mixture, $^{173}$Yb-$^{174}$Yb, where Feshbach resonances are not applicable. Instead, we analyze the tunneling channel by tuning the particle densities and the shape of the potential barrier. Once the Goldstino tunneling channel is identified, it could provide the evidence for the existence of supersymmetry in such Bose-Fermi mixtures.
\begin{figure}[t]
    \centering
    \includegraphics[width=8cm]{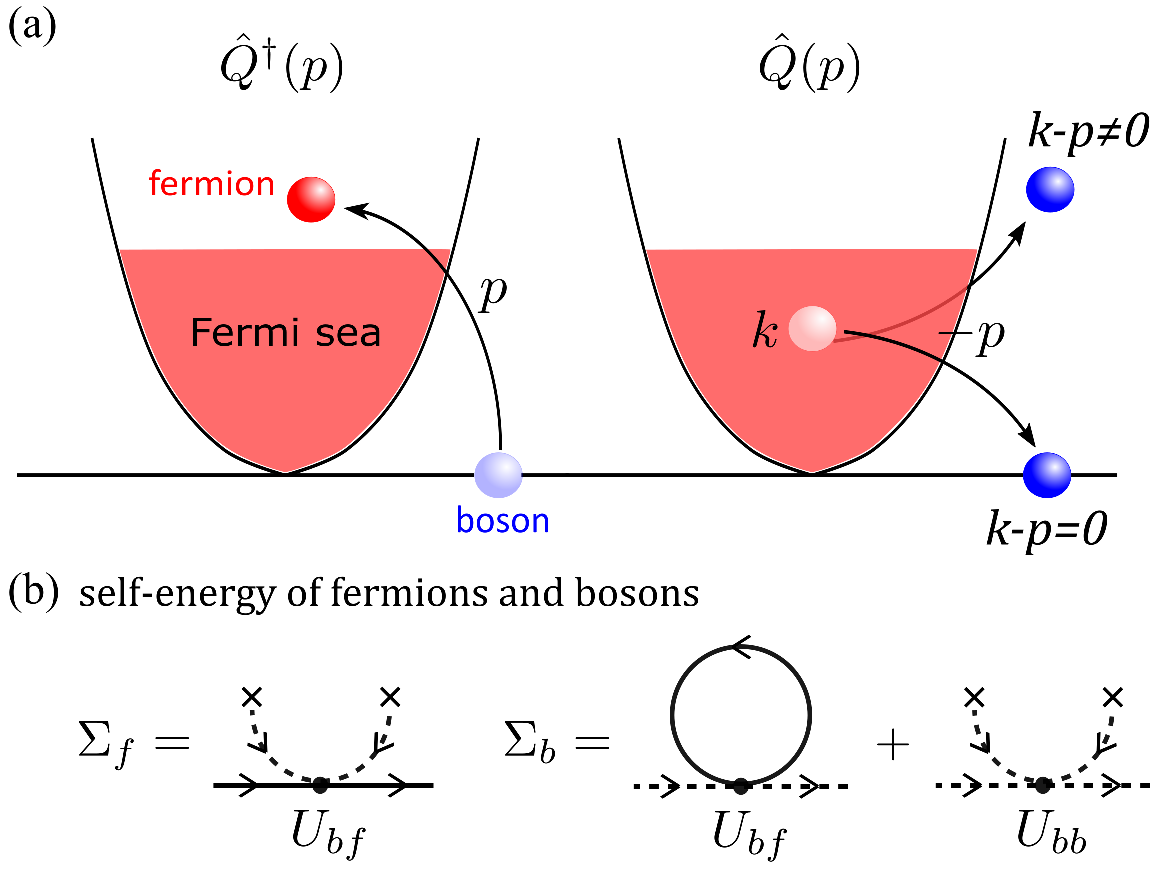}
    \caption{(a) The supercharge operators. The left process represents $\hat{Q}^\dagger$, which converts a boson into a fermion, while the right one represents $\hat{Q}$, converting a fermion into a boson. The black line at the bottom represents the condensate of bosons. (b) Diagrammatic presentation of leading-order self-energies for fermions ($\Sigma_f$) and bosons ($\Sigma_b$). The cross connected to one end of a dashed line denotes the contribution of the condensate of bosons.
    }\label{supercharge}
\end{figure}

This paper is structured as follows: In Sec.~\ref{Formalism}, we introduce the Hamiltonian of a two terminal model. 
In Sec.~\ref{SUSY current}, we derived the formulas of the supersymmetry-like tunneling current and shot noise up to leading order. 
In Sec.~\ref{Fano factor}, we analyze the Fano factor for supersymmetry-like transport and and show the numerical results with changing particle densities and barrier potentials. 
The conclusion and perspectives are given in Sec.~\ref{Summary}.

Throughout the paper, we take $\hbar=k_{\rm B}=1$ and the volumes of both reservoirs to be unity.

\section{Formalism}\label{Formalism}

We consider a two-terminal model for mixtures of ultracold Bose and Fermi gases, where particles in one reservoir have a probability of tunneling through a barrier potential to the other side, as depicted in Fig.~\ref{schematic} (a). In this setup, the bosons predominantly exist in the BEC phase with zero momentum below the BEC critical temperature, while the fermions occupy a fully filled Fermi sea.  We assume the gas is homogeneous within each reservoir, far from the barrier, allowing the wave functions to be asymptotically treated as plane waves. Then the effective Hamiltonian of the total system can be given as $\hat{H}=\hat{H}_{\rm L}+\hat{H}_{\rm R}+\hat{H}_{\rm 1t}+\hat{H}_{\rm 2t}$~\cite{PhysRevB.110.064512}, where the reservoir Hamiltonian reads 
\begin{align}
    \hat{H}_{i={\rm L, R}}=\,&
\sum_{\bm{k}} 
\varepsilon_{\bm{k}, f} 
\hat{f}_{\bm{k}, i}^{\dagger} 
\hat{f}_{\bm{k}, i} 
+\sum_{\bm{k}} 
\varepsilon_{\bm{k}, b} 
\hat{b}_{\bm{k}, i}^{\dagger} \hat{b}_{\bm{k}, i} \nonumber\\
&+\frac{U_{bb}}{2} \sum_{\bm{P}, \bm{q}, \bm{q}'} \hat{b}_{\frac{\bm{P}}{2}+\bm{q}, i}^{\dagger} 
\hat{b}_{\frac{\bm{P}}{2}-\bm{q}, i}^{\dagger} 
\hat{b}_{\frac{\bm{P}}{2}-\bm{q}', i} 
\hat{b}_{\frac{\bm{P}}{2}+\bm{q}', i}\nonumber\\
&+U_{bf} \sum_{\bm{P}, \bm{q}, \bm{q}'} \hat{b}_{\frac{\bm{P}}{2}+\bm{q}, i}^{\dagger} 
\hat{f}_{\frac{\bm{P}}{2}-\bm{q}, i}^{\dagger} 
\hat{f}_{\frac{\bm{P}}{2}-\bm{q}', i} 
\hat{b}_{\frac{\bm{P}}{2}+\bm{q}', i}.
\end{align}
Here $\hat{f}(\hat{b})$ denotes the fermionic (bosonic) annihilation operator, and $\varepsilon_{\bm{p},\alpha=b,f}=p^2/(2m_{\alpha})$ is the kinetic energy of particles. $U_{bb}$ and $U_{bf}$ are, respectively, the coupling strengths for the boson-boson and boson-fermion interactions, which are characterized by the scattering lengths $a_{bb}$ and $a_{bf}$ as $U_{bb}=(4\pi a_{bb})/m_b$ and $U_{bf}=(2\pi a_{bf})/m_r$, with $m_r=1/(1/m_b+1/m_f)$ denoting the reduced mass. In the followings we consider a mixture with small mass imbalance, where $m_f\approx m_b\approx m$. Notice that the fermion-fermion interaction disappears since we consider a polarized Fermi gas. For the tunneling Hamiltonian, we divide it into one-body ($\hat{H}_{\rm 1t}$) and two-body ($\hat{H}_{\rm 2t}$) terms, where the one-body term is given by
\begin{align}
    \hat{H}_{\rm 1t}=&\sum_{\bm{p},\bm{q}}\mathcal{T}_{f,\bm{p},\bm{q}}\hat{f}^\dagger_{\bm{p},{\rm L}}\hat{f}_{\bm{q},{\rm R}}+{\rm H.c.}\cr
    &+\sum_{\bm{p},\bm{q}}\mathcal{T}_{b,\bm{p},\bm{q}}\hat{b}^\dagger_{\bm{p},{\rm L}}\hat{b}_{\bm{q},{\rm R}}+{\rm H.c.}.
\end{align}
We apply rectangular potential barriers with width $\lambda$ and height $V_\alpha$ ($V_f$ for fermions and $V_b$ for bosons) as shown in Fig~\ref{schematic} (b). The one-body tunneling amplitude $\mathcal{T}_{\alpha,\bm{p},\bm{q}}$ can be given by (see Appendix~\ref{appendixA})
\begin{align}
    \mathcal{T}_{\alpha,\bm{p},\bm{q}}\simeq T_{\alpha,\bm{q}}\delta_{p_z,q_z}
    \varepsilon_{q}+V_\alpha\mathcal{C}_{\alpha,p_z,q_z},
\end{align}
where we omit the Hartree terms $\frac{U_{fb}}{2}\sum_i \hat{N}_{b,\bm{p}-\bm{q},i}$, $\frac{U_{bb}}{2}\sum_i \hat{N}_{b,\bm{p}-\bm{q},i}$, and $\frac{U_{fb}}{2}\sum_i \hat{N}_{f,\bm{p}-\bm{q},i}$, which are negligible due to the short-range interaction.
Here $p_z$ is the $z$ component of momentum $\bm{p}$, $T_{\alpha,\bm{p}}$ is the single-particle transmission coefficient through the barrier, and $\mathcal{C}_{\alpha,p_z,q_z}$ represents the overlap of wave functions within the potential barrier. Then by assuming the momentum conservation during the tunneling process ($\bm{p}-\bm{q}\rightarrow0$), the one-body tunneling strength becomes $\mathcal{T}_{\alpha,\bm{p},\bm{q}}=\mathcal{T}_{\alpha,\bm{p}}\delta_{\bm{p}\bm{q}}=(T_{\alpha,\bm{p}}\varepsilon_{p}+V_\alpha\mathcal{C}_{\alpha,p_z})\delta_{\bm{p}\bm{q}}$.
\begin{figure*}[t]
    \centering
    \includegraphics[width=15cm]{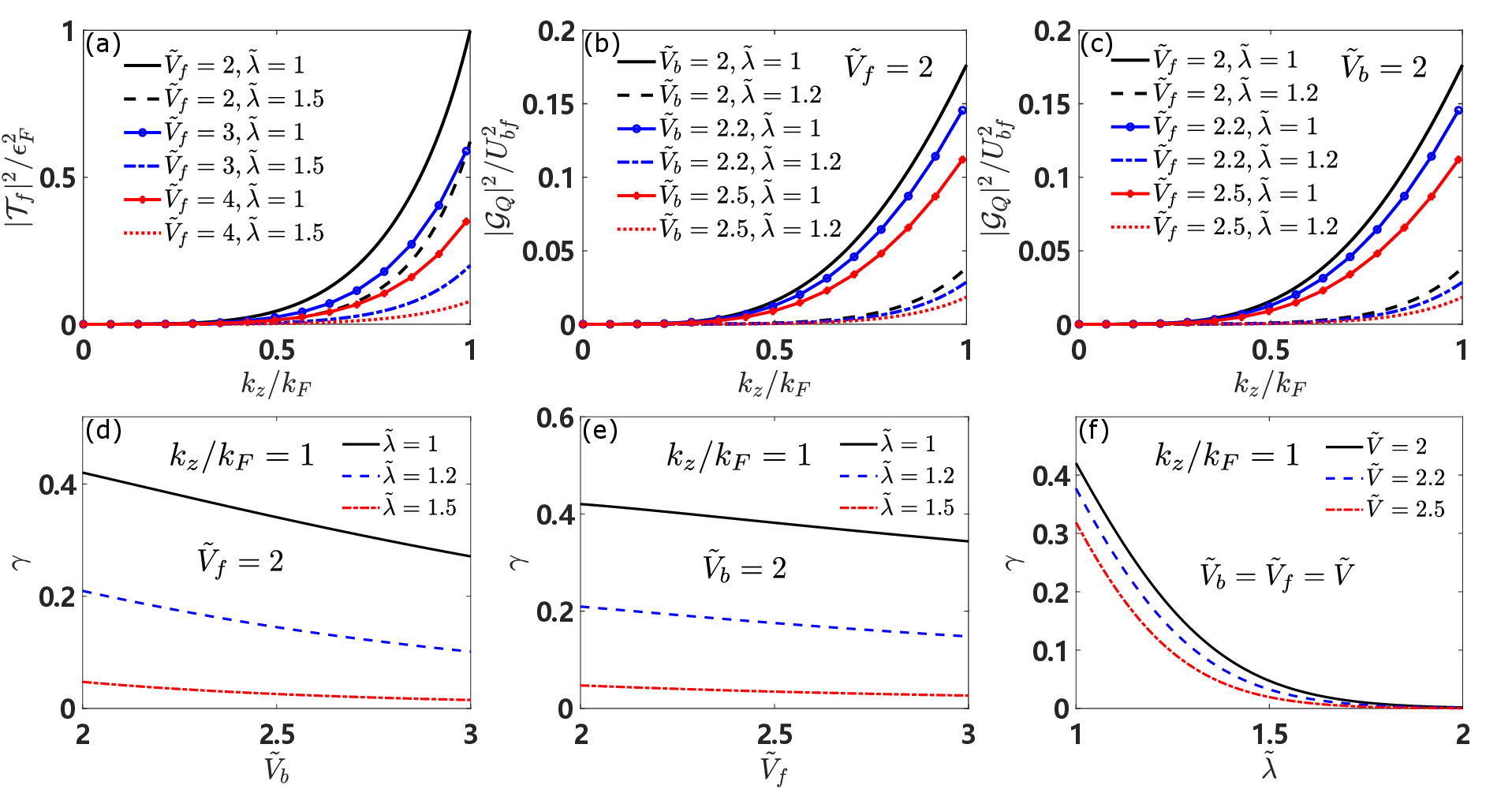}
    \caption{Upper panels: Momentum dependence of (a) module squared fermionic quasiparticle tunneling strengths, (b) module squared supercharge tunneling strengths with different $\tilde{V}_b$, and (c) module squared supercharge tunneling strengths with different $\tilde{V}_f$. lower panels: values of $\gamma=\epsilon_{\rm F}\mathcal{G}_Q/(U_{bf}\mathcal{T}_f)$ with changing (d) height of bosonic barrier, (e) height of fermionic barrier, and (f) width of barriers. Here $\tilde{V}_\alpha=V_\alpha/\epsilon_{\rm F}$ and $\tilde{\lambda}=\lambda k_{\rm F}$.
    }\label{strengths}
\end{figure*}

The two-body tunneling Hamiltonian includes three parts: $\hat{H}_{\rm 2t}=\hat{H}_{bb}+\hat{H}_{bf}+\hat{H}_{Q}$, respectively corresponding to contributions of boson-boson pairs, boson-fermion pairs and the supercharges. By defining the annihilation operators for boson-boson and boson-fermion pairs: $\hat{P}_{bb,i}(\bm{p})=\sum_{\bm{k}}\hat{b}_{-\bm{k}+\bm{p}/2,i}\hat{b}_{\bm{k}+\bm{p}/2,i}$ and $\hat{P}_{bf,i}(\bm{p})=\sum_{\bm{k}}\hat{b}_{-\bm{k}+\bm{p}/2,i}\hat{f}_{\bm{k}+\bm{p}/2,i}$, the pair tunneling terms can be written as 
\begin{subequations}
    \begin{align}
        \hat{H}_{bb}=\,&\frac{1}{2}\mathcal\sum_{\bm{p},\bm{q}}{G}_{bb,\bm{p},\bm{q}}\hat{P}^\dagger_{bb, {\rm L}}(\bm{p})\hat{P}_{bb, {\rm R}}(\bm{q})+\textrm{H.c.},
    \end{align}
    \begin{align}
        \hat{H}_{bf}=\,&\mathcal\sum_{\bm{p},\bm{q}} {G}_{bf,\bm{p},\bm{q}}\hat{P}^\dagger_{bf, {\rm L}}(\bm{p})\hat{P}_{bf, {\rm R}}(\bm{q})+\textrm{H.c.},
    \end{align}
\end{subequations}
The pair tunneling amplitude $\mathcal{G}_{bb}$ and $\mathcal{G}_{bf}$ can be expressed as $\mathcal{G}_{bb,\bm{p},\bm{q}}=\frac{1}{2}U_{bb}(T^*_{b,\bm{p}}T^*_{b,-\bm{p}}+T_{b,-\bm{q}}T_{b,\bm{q}})$ and $\mathcal{G}_{bf,\bm{p},\bm{q}}=\frac{1}{2}U_{bf}(T^*_{f,\bm{p}}T^*_{b,-\bm{p}}+T_{b,-\bm{q}}T_{f,\bm{q}})$.
The supercharge tunneling term is given by
\begin{align}
    \hat{H}_{Q}=\,\sum_{\bm{p},\bm{q}}\mathcal{G}_{Q,\bm{p},\bm{q}}\left[\hat{Q}_{\rm L}^\dagger(\bm{p})\hat{Q}_{\rm R}(\bm{q})+\textrm{H.c.}
\right].
\end{align}
Here $\hat{Q}$ is the supercharge annihilation operator defined by
\begin{align}
    \hat{Q}_i(\bm{p})=\sum_{\bm{k}}\hat{f}_{\bm{k},i}\hat{b}^\dagger_{\bm{k}-\bm{p},i},
\end{align}
which satisfy the anticommutation relation and therefore is a fermionic operator.
Physically this operator annihilates a fermion while creating a boson with momentum transferred, as shown in Fig.~\ref{supercharge} (a). The tunneling amplitude $\mathcal{G}_Q$ is given by $\mathcal{G}_{Q,\bm{p},\bm{q}}=\frac{1}{2}U_{bf}(T_{b,\bm{p}}T^*_{f,\bm{p}}+T_{f,\bm{q}}T^*_{b,\bm{q}})$. Again with the assumption of momentum conservation, we have $\mathcal{G}_{bb,\bm{p},\bm{q}}=\mathcal{G}_{bb,\bm{p}}\delta_{\bm{p}\bm{q}}=U_{bb}\operatorname{Re}[T^*_{b,\bm{p}}T^*_{b,-\bm{p}}]\delta_{\bm{p}\bm{q}}$, $\mathcal{G}_{bf,\bm{p},\bm{q}}=\mathcal{G}_{bf,\bm{p}}\delta_{\bm{p}\bm{q}}=U_{bf}\operatorname{Re}[T^*_{f,\bm{p}}T^*_{b,-\bm{p}}]\delta_{\bm{p}\bm{q}}$, and $\mathcal{G}_{Q,\bm{p},\bm{q}}=\mathcal{G}_{Q,\bm{p}}\delta_{\bm{p}\bm{q}}=U_{bf}\operatorname{Re}[T_{b,\bm{p}}T^*_{f,\bm{p}}]\delta_{\bm{p}\bm{q}}$.

The supersymmetry-like current operator yields $\hat{I}_{\rm SUSY}=i\big[\hat{N}_{b,{\rm L}}-\hat{N}_{f,{\rm L}},\hat{H}\big]$, where $\hat{N}_{f,i}=\sum_{\bm{k}}\hat{f}^\dagger_{\bm{k},i}\hat{f}_{\bm{k},i}$ and $\hat{N}_{b,i}=\sum_{\bm{k}}\hat{b}^\dagger_{\bm{k},i}\hat{b}_{\bm{k},i}$ represent particle number densities of fermions and bosons, respectively. 
In order to facilitate the identification of Goldstino current, we consider the same bosonic chemical potential but different fermionic chemical potentials in the two reservoirs, such that the supercharge tunneling can be induced while the bosonic quasiparticle and boson-boson pair tunneling do not occur. On the other hand, we consider a short-range attractive interactions with positive scattering length, where the quantum gas encounters an instability toward a ground state with pair formation in the strong coupling regime~\cite{Massignan_2014,PhysRevLett.129.203402}. In addition, this process would compete with the phase separation occurring in a strong-interacting Bose-Fermi mixture, where effective repulsive interaction is originated from the attractive interaction.
Nevertheless, to avoid this we consider the weak coupling regime, where the two-particle bound state is significantly deep in the energy spectrum and the Bose-Fermi mixture remains at a metastable state. In this case, our approach is valid while the boson-fermion pair tunneling process is suppressed. Even if we consider the contribution of boson-fermion pairs, the distribution functions included in the formula of tunneling current, $f_f(\omega-\Delta\mu)-f_f(\omega)$, would make it very small, since the energy of the pairs is far below zero. Thus the remaining contributions come from the fermionic quasiparticle and supercharge, whose current operators read 
\begin{align}
    \hat{I}_{\rm 1t}=\,i\sum_{\bm{p}, \bm{q}} \mathcal{T}_{f,\bm{p},\bm{q}}\hat{f}_{\bm{p},  {\rm R}}^{\dagger} \hat{f}_{\bm{q}, {\rm L}}+{\rm H.c.},
\end{align}
\begin{align}
    \hat{I}_{Q}=\,2i\sum_{\bm{p},\bm{q}}\mathcal{G}_{Q,\bm{p},\bm{q}}\hat{Q}_{\rm L}^\dagger(\bm{p})\hat{Q}_{\rm R}(\bm{q})+\textrm{H.c.}.
\end{align}
The momentum dependence of their tunneling strengths is depicted in Figs.~\ref{strengths} (a), (b), and (c), where $\epsilon_{\rm F}$ and $k_{\rm F}$ are Fermi energy and Fermi momentum of the Fermi gas in the left reservoir. It is shown that $\mathcal{G}_Q$ is more sensitive than $\mathcal{T}_f$ to the shape of potential barrier. To compare the two tunneling strengths more intuitively, we define the ratio $\gamma=\epsilon_{\rm F}\mathcal{G}_Q/(U_{bf}\mathcal{T}_f)$. The tunneling of particles near the Fermi surface with large momentum $k_z$ plays a dominant role in the total tunneling process, making it reasonable to approximately regard the tunneling strength as constants 
$\mathcal{T}_f\equiv \mathcal{T}_{f,k_z=k_{\rm F}}$ and $\mathcal{G}_Q=\mathcal{G}_{Q,k_z=k_{\rm F}}$. Figs.~\ref{strengths} (d), (e), and (f) exhibit how this ratio is affected by the height and width of the potential barrier. The details of calculation can be checked in Appendix~\ref{appendixA}.

\section{Supersymmetry-like current and noise}\label{SUSY current}

By expanding the formulas of tunneling currents up to the leading order within the Schwinger-Keldysh Green's function approach~\cite{Schwinger,Keldysh}, and using the Langreth rule, we obtain the leading-order formulas of the currents:
\begin{align}\label{I1t}
    I_{\rm 1t}=\,4 |\mathcal{T}_f|^2 \int \frac{d \omega}{2 \pi} &\sum_{\bm{p}, \bm{q}}
        %\mathcal{T}_{\bm{p}, \bm{q}, f} \mathcal{T}_{\bm{q}, \bm{p}, f}
        \operatorname{Im} G_{f, \bm{p}, \mathrm{L}}(\omega-\Delta \mu_f) \operatorname{Im} G_{f, \bm{q}, \mathrm{R}}(\omega)\nonumber\\
        &\times[f_f(\omega-\Delta \mu_f)-f_f(\omega)],
\end{align}
\begin{align}\label{IQ}
    I_{Q}=8|\mathcal{G}_{Q}|^{2} \sum_{\bm{p},\bm{q}} \int \frac{d \Omega}{2 \pi} &\operatorname{Im} \chi_{\bm{p}, {\rm L}}(\Omega-\Delta\mu_{Q}) \operatorname{Im} \chi_{\bm{q}, {\rm R}}(\Omega)\nonumber\\
    \times&[f_f(\Omega-\Delta\mu_{Q})-f_f(\Omega)],
\end{align}
\begin{figure}[t]
    \centering
    \includegraphics[width=8.6cm]{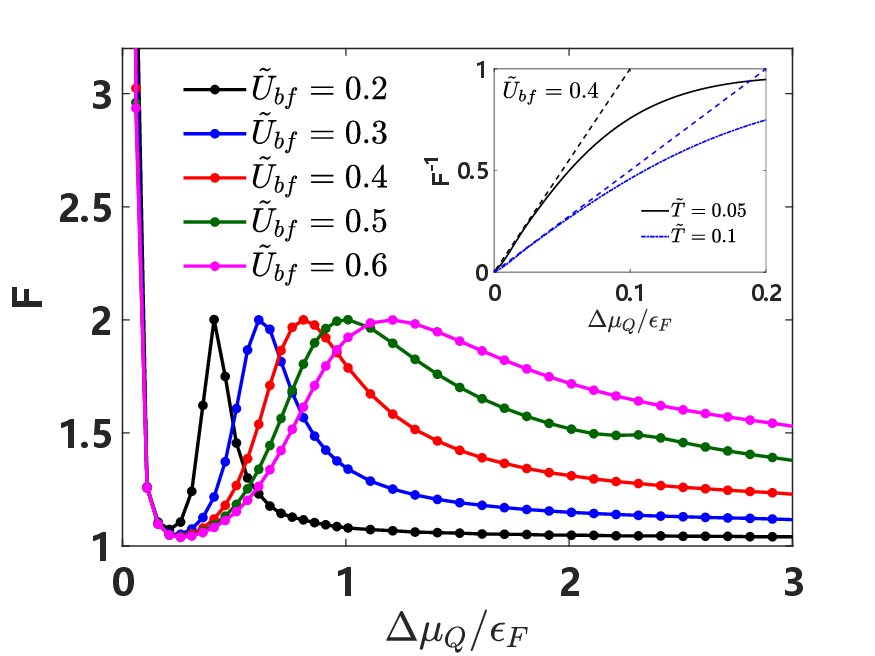}
    \caption{
    Bias dependence of the Fano factor for the supersymmetry-like 
    transport in a $^{173}$Yb-$^{174}$Yb ultracold mixture, where $\tilde{U}_{bf}=N_f U_{bf}/\epsilon_{\rm F}=4k_{\rm F}a_{bf}/(3\pi)$. We take 
    $\gamma=0.25$. The temperature is taken to be $\tilde{T}=0.05$, where we define $\tilde{T}=T/T_{\rm F}$, with $T_{\rm F}$ denoting the Fermi temperature. Ratio between bosonic and fermionic densities is set to be $N_b/N_f=2$. The Onsager's relation $F^{-1}(\Delta \mu_Q\rightarrow 0)=\Delta \mu_Q/2T$ is shown in the inset.
    }\label{F-d}
\end{figure}
where $f_f(\omega)=1/e^{\omega/T}+1$ is the Fermi distribution function.
With $\mu_{b(f),i}$ denoting the chemical potential of bonsons (fermions) in reservoir $i$, we define $\Delta\mu_f=\mu_{f,{\rm L}}-\mu_{f,{\rm R}}$ and $\Delta\mu_{Q}=\mu_{Q,{\rm L}}-\mu_{Q,{\rm R}}$ as the chemical potential bias of fermions and supercharge, where $\mu_{Q,i}=\mu_{f,i}-\mu_{b,i}$. $G_{f,\bm{p},i}$ is the retarded single-particle Green's function and $\chi$ is the retarded Goldstino propagator, which is defined as $\chi_{\bm{p}}(t,t')=i\theta(t-t')\langle\{\hat{Q}_{\bm{p}}(t),\hat{Q}^\dagger_{\bm{p}}(t')\}\rangle$ in the time representation. Notice that although the Goldstino tunneling current $I_Q$ is a kind of boson-fermion tunneling perturbation. However, different from the boson-fermion pair process, the boson and fermion operators couple with Goldstino propagators rather than boson-fermion pair propagators, which read $\Gamma_{bf,\bm{p},i}(t,t')=-i\theta(t-t')\langle\{\hat{P}_{bf,\bm{p},i}(t)\hat{P}^\dagger_{bf,\bm{p},i}(t')\}\rangle$. The Goldstino propagator can be calculated, by using the random phase approximation (RPA)~\cite{PhysRevA.96.063617,PhysRevResearch.3.013035,PhysRevB.110.064512}, as 
\begin{align}\label{RPA}
    \chi_{\bm{p}}(\Omega)=\frac{\Pi_{\bm{p}}(\Omega)}{1+U_{bf}\Pi_{\bm{p}}(\Omega)},
\end{align}
with $\Pi_{\bm{p}}(\Omega)$ denoting the one-loop diagram, namely, the non-interacting Goldstino propagator. We can see from Fig.~\ref{supercharge} that $\Pi_{\bm{p}}(\Omega)$ includes the contribution of both the condensate and the bosons with non-zero momentum. We can thus write $\Pi_{\bm{p}}$ as the sum of two parts: $\Pi_{\bm{p}}=\Pi^{\rm p}_{\bm{p}}+\Pi^{\rm c}_{\bm{p}}$, with 
\begin{align}\label{Pip}
    \Pi^{\rm p}_{\bm{p}}(\Omega)=-\frac{N^0_b}{\Omega+\epsilon_{\rm F}+i\delta-\bm{p}^2/2m},
\end{align}
\begin{align}\label{Pic}
    \Pi^{\rm c}_{\bm{p}}(\Omega)=-\int\frac{d^3\bm{k}}{(2\pi)^3}
    \frac{{f}_f(\xi_{\bm{k}+\bm{p},f})+{f}_b(\xi_{\bm{k},b})}{\Omega+i\delta-\xi_{\bm{k}+\bm{p},f}+\xi_{\bm{k},b}},
\end{align}
where $N_{b}^0$ denotes the number of bosons in the condensate state, and $f_b(\omega)$ denotes the distribution function of bosons out of the condensate. (The calculation details of the Goldstino propagator and feature of its spectrum are shown in Appendix~\ref{appendixB}.) We define $\xi_{\bm{k},f}=\bm{k}^2/(2m_f)-\mu_f+\Sigma_{f}$ and $\xi_{\bm{k},b}=\bm{k}^2/(2m_b)-\mu_b+\Sigma_{b}$, where $\Sigma_f$ and $\Sigma_b$ denote the mean-field corrections for the single-particle energies of fermions and bosons. According to Fig.~\ref{supercharge} (b), for fermions and bosons with nonzero momentum, the self-energies respectively read $\Sigma_f=U_{bf}N_b$ and $\Sigma_{b}=2U_{bb}N_b+U_{bf}N_f$, while for condensate bosons it turns out to be $\Sigma_{b,0}=U_{bb}N_b+U_{bf}N_f$ due to the absence of an exchange term~\cite{PhysRevA.96.063617}.
Eq.~(\ref{Pip}) is the contribution of condensate bosons, leading to a series of poles at $\Omega+\epsilon_{\rm F}-\bm{p}^2/2m=0$ in the Goldstino spectrum. In contrast, Eq.~(\ref{Pic}) arises from the boson with non-zero momentum, resulting in the continuum of the spectrum~\cite{PhysRevA.96.063617,PhysRevB.110.064512}. At extremely low temperatures, almost all bosons are distributed in the BEC condensate, implying $N_b^0\rightarrow N_b$ and $f_b(\xi_{\bm{k},b})\rightarrow 0$. Apart from the mean-field correction, an additional modification for the one-loop propagator comes from the Bogoliubov excitations at low-momentum region. However, here we neglect this modification term since it is much smaller than the contribution given by Eq.~(\ref{Pic}) (see Appendix~\ref{appendixC}). 

Now we introduce the current shot noise , which is defined as $\hat{\mathcal{S}}(t_1,t_2)=\frac{1}{2}\langle\hat{I}(t_1)\hat{I}(t_2)+\hat{I}(t_2)
\hat{I}(t_1)\rangle$~\cite{PhysRevB.46.12485,lumbroso2018electronic}, where the quantum average $\langle\cdots\rangle$ represents quantum average with respect to the unperturbed state of the system. For the supersymmetry-like current we have $\hat{\mathcal{S}}_{\rm SUSY}=\hat{\mathcal{S}}_{\rm 1t}+\hat{\mathcal{S}}_{Q}$, corresponding to the one-body and Goldstino transport processes, respectively. Its Fourier decomposition reads
\begin{align}
    \mathcal{S}(\omega)=\frac{1}{\tau}\int_{0}^{\tau}dt_1\int_{0}^{\tau}dt_2
    e^{i\omega(t_1-t_2)}\mathcal{S}(t_1,t_2),
\end{align}
where $\tau$ is the typical time scale for the noise measurement and should be much larger than the transport time scale~\cite{science.1242308}. Here we simply take the limit $\tau\rightarrow\infty$ and the dc-limit of noise: $\mathcal{S}:=\mathcal{S}(\omega\rightarrow 0)$. With the truncation up to the first order, we obtain the following expressions:
\begin{align}\label{S1t}
    \mathcal{S}_{\rm 1t}=&4|\mathcal{T}_f|^2\sum_{\bm{p},\bm{q}}\int \frac{d\omega}{2\pi}
    \operatorname{Im}G_{f,\bm{q},{\rm L}}(\omega-\Delta\mu_f)
    \operatorname{Im}G_{f,\bm{p},{\rm R}}(\omega)\cr
    \times\big[f_f&(\omega-\Delta\mu_f)
    (1-f_f(\omega))+(1-f_f(\omega-\Delta\mu_f))f_f(\omega)\big],
\end{align}
\begin{align}\label{SQ}
    \mathcal{S}_Q=&16|\mathcal{G}_Q|^2\sum_{\bm{p},\bm{q}}
    \int\frac{d\omega}{2\pi}\operatorname{Im}\chi_{\bm{p},{L}}(\omega-\Delta\mu_Q)
    \operatorname{Im}\chi_{\bm{q},{R}}(\omega)\cr
    \times\big[f_f&(\omega-\Delta\mu_Q)\big(1-f_f(\omega)\big)+
    \big(1-f_f(\omega-\Delta\mu_Q)\big)f_f(\omega)\big].
\end{align}
Compared Eqs.~(\ref{S1t}), (\ref{SQ}) with Eqs.~(\ref{I1t}), (\ref{IQ}), one can easily find that at large bias limit ($\Delta\mu_f\rightarrow \infty$ and $\Delta\mu_Q\rightarrow \infty$), we have $\mathcal{S}_{\rm 1t}/I_{\rm 1t}=1$ and $\mathcal{S}_{Q}/I_{Q}=2$. This indicate that the value of the ratio $(\mathcal{S}_{\rm 1t}+\mathcal{S}_{Q})/(I_{\rm 1t}+I_{Q})$ lies between $1$ and $2$, and can reflect the ratio of two types of charge carriers.

\section{Fano factor}\label{Fano factor}

To investigate the tunneling channel of the supersymmertry-like transport, we define the Fano factor as $F=(\mathcal{S}_{\rm 1t}+\mathcal{S}_{Q})/(I_{\rm 1t}+I_{Q})$, which is a useful probe for the current carrier. We consider the $^{173}$Yb-$^{174}$Yb mixture~\cite{PhysRevA.79.021601,sugawa2011interaction} as the context, where the boson-fermion and boson-boson scattering lengths are experimentally determined as $a_{bf}=138.49 a_0$ and $a_{bb}=104.72 a_0$~\cite{PhysRevA.82.011608} ($a_0$ is the Bohr radius). By setting the Fermi energy and the density of fermions of the left reservoir to $\epsilon_{\rm F}$ and $N_f$, and tuning the chemical potential of fermions in the right side, we analyze The bias dependence of the Fano factor as shown in Fig.~\ref{F-d}, where we define the dimensionless interaction strength $\tilde{U}_{bf}=N_f U_{bf}/\epsilon_{\rm F}=4k_{\rm F}a_{bf}/(3\pi)$, and fix the ratio of densities of two components as $N_b/N_f=2$. We note that although the scattering lengths can not be adjusted through magnetic Feshbach resonances, we can tune the $\tilde{U}_{bf}$ by changing the $k_{\rm F}$, namely the particle density of fermions.
\begin{figure}[t]
    \centering
    \includegraphics[width=8.6cm]{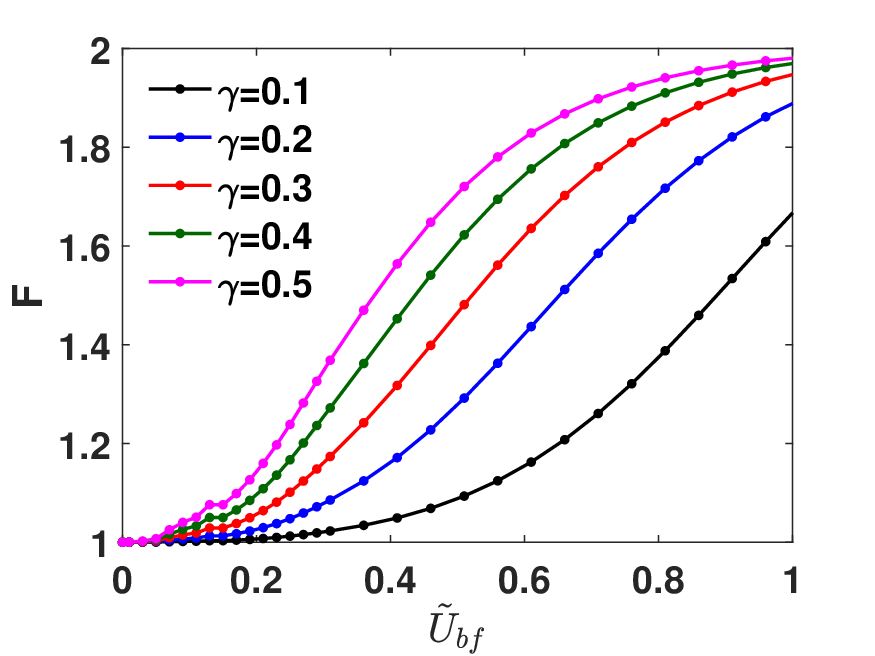}
    \caption{
    Interaction dependence of the Fano factor for the supersymmetry-like 
    transport with different values of $\gamma$ in a $^{173}$Yb-$^{174}$Yb ultracold mixture. We take $\Delta\mu_Q/\epsilon_{\rm F}=3$ to maintain the large bias limit. The temperature is taken to be $T/T_{\rm F}=0.05$, and ratio between bosonic and fermionic densities is set to be $N_b/N_f=2$.
    }\label{F-U}
\end{figure}
Fig.~\ref{F-d} demonstrates the validity of large bias limit when $\Delta\mu_Q$ reaches a relatively large value, such as $\Delta\mu_Q/\epsilon_{\rm F}=3$. In addition, a divergence appears at $\Delta \mu_Q\rightarrow 0$, resulting from zero current but finite noise. The inset figure shows the Onsager's relation near $\Delta \mu_Q=0$, which yields $F^{-1}=\Delta \mu_Q/2T$.

Fig.~\ref{F-U} shows the Fano factor as a function of the dimensionless interaction strength $\tilde{U}_{bf}$ in the $^{173}$Yb-$^{174}$Yb ultracold mixture where particle densities are varied while keeping the ratio $N_b/N_f=2$ fixed. We focus on the weakly coupling regime to avoid the unstability toward pair formation~\cite{Massignan_2014,PhysRevLett.129.203402} and phase separation~\cite{PhysRevLett.121.253602} in the strongly interacting regime . We set $\Delta\mu_Q/\epsilon_{\rm F}=3$ for satisfying the large-bias limit, which can be justified according to Fig.~\ref{F-d}. We see that the Fano factor gradually changes from $1$ to a value close to $2$ with increasing $\tilde{U}_{bf}$, indicating a crossover from fermionic quasiparticle tunneling ($F=1$) to Goldstino tunneling ($F=2$). Similar phenomenon has also been found in the BCS-BEC crossover regime~\cite{pgad045} and repulsively interacting Fermi gases~\cite{PhysRevApplied.21.L031001}, where the current carrier changes from quasiparticles to pairs and magnon modes, respectively. It is also worth noting that the interaction dependence of $F$ is deeply related to the value of $\gamma$, which is characterizes by the property of the junction barrier, which indicates that the tunneling channel can be tuned by adjusting the potential barrier. 

Although the results in this work are specified for the $^{173}$Yb-$^{174}$Yb superfluid mixture, our approach is applicable for other Bose-Fermi mixtures with small mass imbalances, including $^6$Li-$^7$Li~\cite{science.1255380,PhysRevLett.118.103403}, $^{39}$K-$^{40}$K~\cite{PhysRevA.78.012503}, $^{40}$K-$^{41}$K~\cite{PhysRevA.84.011601}, $^{84}$Sr-$^{87}$Sr~\cite{PhysRevA.82.011608}, $^{87}$Rb-$^{87}$Sr~\cite{barbe2018observation}, and $^{161}$Dy-$^{162}$Dy~\cite{PhysRevLett.103.085301}. Moreover, Feshbach resonances allow tuning the interspecies scattering lengths in some of these mixtures~\cite{PhysRevA.84.011601,barbe2018observation}, offering a more flexible experimental platform to study the effects of interspecies interactions on the tunneling channels and Goldstino spectral properties. A noise measurement experiment for density current in cold atoms has been proposed with a two-terminal system~\cite{PhysRevA.98.063619}, where one reservoir overlaps with a single-sided optical cavity, and a probe beam monitors the atom number in this reservoir. The current noise can be measured through the phase of the probe, which gives a real-time measurement of the atom number. Such method might also be applicable for the supersymetry-like tunneling in the Bose-Fermi mixture, where the number of bosons and fermions should be measured respectively.

\section{Conclusion}\label{Summary}

In this study, we theoretically investigate the tunneling transport induced by chemical potential bias and the corresponding shot noise in an ultracold Bose-Fermi mixture with explicit supersymmetry breaking. We show how the shot noise can serve as a probe of the current carrier at large bias limit. The Fano factor, defined as the noise-to-current ratio, takes the value $1$ at noninteracting case, indicating quasiparticles dominant the tunneling process. As the density of atoms increases, the Fano factor changes to a value close to $2$, signifying a transition to Goldstino-dominated tunneling. Furthermore, we show that the tunneling channel can be controlled by adjusting the junction barrier, which could facilitate experimental investigations of supercharge tunneling transport. Our approach offers a novel method for detecting the Goldstino in mixtures of bosonic and fermionic atoms and may also prove useful for exploring Nambu-Goldstone fermions in other condensed matter systems with supersymmetric properties~\cite{PhysRevD.99.045002, marra20221d, miura2023}. 

\begin{acknowledgements}
T. Z. thanks Hiroyuki Tajima and Haozhao Liang for useful discussions.
T. Z. was supported by the RIKEN Junior Research Associate Program.

\end{acknowledgements}

\appendix

\section{Calculation of tunneling coupling strengths}\label{appendixA}

In this appendix, we show the details of our caluculation of tunneling coupling strengths $\mathcal{T}_f$ and $\mathcal{G}_Q$, as well as the ratio $\gamma$.
We consider a rectangular barrier potential:
\begin{align}
    V_\alpha(z)=\left\{
    \begin{array}{cc}
        V_\alpha  & 0 <z < \lambda  \\
        0  &  z\leq 0, \ z\geq \lambda
    \end{array}
    \right.
\end{align}
Here we consider a particle from the left reservior ($z<0$) to the right reservoir 
($z\geq \lambda$) described by 
$\Psi_{\alpha,\bm{k},{\rm L}}(\bm{r})=e^{i\bm{k}_\perp\cdot\bm{r}_\perp}
\psi_{\alpha,k_z,{\rm L}}(z)$ 
where $\bm{r}_\perp=(x,y)$ and $\bm{k}_\perp=(k_x,k_y)$ are the position and 
momentum vectors perpendicular to the barrier, respectively. According to the 
behavior of a wave function passing through a potential barrier, 
$\psi_{k_z,{\rm L}}(z)$ for a particle coming from the left reservoir can be written as 
\begin{align}
    \psi_{\alpha,k_z,{\rm L}}(z)=
    \left\{
    \begin{array}{cc}
      e^{ik_z z}+R_\alpha e^{-ik_z z} 
      &  z\leq 0, \\
        C_{1\alpha}e^{\kappa_{z\alpha} z}+C_{2\alpha}e^{-\kappa_{z\alpha} z} & 
        0<z<\lambda, \\
    T_\alpha e^{ik_z z} 
    & z\geq\lambda.
    \end{array}
    \right.  
\end{align}
where $\kappa_{z\alpha}=\sqrt{2mV_\alpha-k_z^2}$ is a real number (we suppose the potential barrier 
is larger than the particle's kinetic energy). According to the continuity of $\psi(z)$ and $d\psi(z)/dz$ at the boundary of barrier $z=0$ and $z=\lambda$, the reflection coefficient $R$ and transmission coefficient $T$ should be
\begin{align}
    R_{\alpha,\bm{k}}&=\frac{2(\kappa_{z\alpha}^2+k_z^2)\sinh{(\kappa_{z\alpha}
    \lambda})}{(\kappa_{z\alpha}+ik_z)^2e^{-\kappa_{z\alpha}\lambda}
    -(\kappa_{z\alpha}-ik_z)^2e^{\kappa_{z\alpha}\lambda}},\label{Rk}\\
    T_{\alpha,\bm{k}}&=\frac{i4k_z\kappa_{z\alpha}e^{-ik_z\lambda}}{(\kappa_{z\alpha}
    +ik_z)^2e^{-\kappa_{z\alpha}\lambda}-(\kappa_{z\alpha}-ik_z)^2e^{\kappa_{z\alpha}
    \lambda}}\label{Tk},
\end{align}
while the coefficients $C_{1\alpha,\bm{k}}$ and $C_{2\alpha,\bm{k}}$ for the wave 
function inside the potential barrier are obtained as 
\begin{align}
    C_{1\alpha,\bm{k}}&=\frac{i2k_z(\kappa_{z\alpha}+ik_z)e^{-\kappa_{z\alpha}\lambda}}
    {(\kappa_{z\alpha}+ik_z)^2e^{-\kappa_{z\alpha}\lambda}-(\kappa_{z\alpha}-ik_z)^2
    e^{\kappa_{z\alpha}\lambda}},\\
    C_{2\alpha,\bm{k}}&=\frac{i2k_z(\kappa_{z\alpha}-ik_z)e^{\kappa_{z\alpha}
    \lambda}}{(\kappa_{z\alpha}+ik_z)^2e^{-\kappa_{z\alpha}\lambda}-(\kappa_{z\alpha}
    -ik_z)^2e^{\kappa_{z\alpha}\lambda}}.
\end{align}
We may obtain the wave function $\Psi_{\alpha,\bm{k},{\rm R}}(\bm{r})$ of a particle coming from the right reservoir to the left one in a same way:
\begin{align}
    \psi_{\alpha,-k_z,{\rm R}}(z)=
    \left\{
    \begin{array}{cc}
        T'_{\alpha}e^{-ik_z z} 
      &  z\leq 0, \\
        C'_{1\alpha}e^{\kappa_z z}+C'_{2\alpha}e^{-\kappa_z z} & 0<z<\lambda, \\ 
    e^{-ik_z z}+R'_{\alpha}e^{ik_z z} ,
    & z\geq\lambda
    \end{array}
    \right.  
\end{align}
where we have the relations: $R'_{\alpha,-\bm{k}}=e^{-i2k_z\lambda}R_{\alpha,\bm{k}}$, 
$C'_{1\alpha,-\bm{k}}=e^{-(\kappa_z+ik_z)\lambda}C_{2\alpha,\bm{k}}$, 
$C'_{2\alpha,-\bm{k}}=e^{(\kappa_z-ik_z)\lambda}C_{1\alpha,\bm{k}}$, and 
$T'_{\alpha,-\bm{k}}=T_{\alpha,\bm{k}}$.

The total Hamiltonian of the system is given by
\begin{align}\label{Hamiltonian}
\hat{H}= &
\sum_{\alpha=f,b}\int d^3 \bm{r} \hat{\psi}_\alpha^{\dagger}(\bm{r})\left[-\frac{\nabla^2}{2 m_\alpha}+V_\alpha(\bm{r})\right] \hat{\psi}_\alpha(\bm{r})\cr
& +\frac{U_{b b}}{2} \int d^3 \bm{r} \hat{\psi}_b^{\dagger}(\bm{r}) \hat{\psi}_b^{\dagger}(\bm{r}) \hat{\psi}_b(\bm{r})\hat{\psi}_b(\bm{r}) \cr
& +U_{b f} \int d^3 \bm{r} \hat{\psi}_b^{\dagger}(\bm{r}) \hat{\psi}_b(\bm{r}) \hat{\psi}_f^{\dagger}(\bm{r}) \hat{\psi}_f(\bm{r}),
\end{align}
For the steady-state transport between two reservoirs,
the field operator can be decomposed as
\begin{align}\label{decomposition}
    \hat{\psi}_{\alpha}(\bm{r})=\hat{\psi}_{\alpha,{\rm L}}(\bm{r})+\hat{\psi}_{\alpha,{\rm R}}(\bm{r}).
\end{align}
We expand the operator $\hat{\Psi}_\alpha,i(r)$ with respect to the asymptotic wave
functions as
\begin{align}
    \hat{\psi}_{f,i}(\bm{r})=\sum_{\bm{k}}\psi_{\bm{k},f,i}(\bm{r})\hat{f}_{\bm{k},i},
\end{align}
\begin{align}
    \hat{\psi}_{b,i}(\bm{r})=\sum_{\bm{k}}\psi_{\bm{k},b,i}(\bm{r})\hat{b}_{\bm{k},i},
\end{align}
where we only use wave functions $z < 0$ and $z > \lambda$,
since we characterize the tunneling strength by using
the asymptotic wave functions far away from the tunnel barrier. Substituting them to the total Hamiltonian and supposing $\bm{k}_\perp$ of the particle 
before and after tunneling to be unchanged, we can write the momentum-dependent 
tunneling strength as
\begin{align}
    \mathcal{T}_{f,\bm{k},\bm{k}'}\simeq T_{f,\bm{k}'}\delta_{k_z,k'_z}
    \varepsilon_{k'}+V_f\mathcal{C}_{f,k_z,k'_z},
\end{align}
where $\varepsilon_{k}=\bm{k}^2/2m$.
For $\mathcal{T}_{f,\bm{k},\bm{k}'}$ we omit the Hartree term 
$\frac{U_{fb}}{2}\sum_i \hat{N}_{b,\bm{k}_1-\bm{k}_2,i}$ which is negligible due to 
the short-range interaction, while
\begin{align}\label{C}
    \mathcal{C}_{f,k_z,k'_z}=\frac{1}{\lambda}\int_0^\lambda dz 
    \psi^\dagger_{f,k_z,{\rm R}}(z)\psi_{f,k'_z,{\rm L}}(z).
\end{align}
Here $\epsilon_{k}=k^2/(2m)$ denotes the particle's kinetic energy.
Now we assume that $\bm{k}-\bm{k}'\rightarrow 0$, and Eq.~(\ref{C}) can be 
approximately written as 
\begin{widetext}
    \begin{align}
    \mathcal{C}_{f,k_z}\simeq&-\frac{\sinh(\kappa_{zf}\lambda)}{\lambda\kappa_{zf}}
    \frac{4k_z^2(\kappa_{zf}^2+k_z^2)e^{-ik_z\lambda}}{[(\kappa_{zf}+ik_z)^2
    e^{-\kappa_{zf}\lambda}-(\kappa_{zf}-ik_z)^2e^{\kappa_{zf}\lambda}]^2}\cr
    &-\frac{8k_z^2e^{-ik_z\lambda}[(\kappa^2_{zf}-k_z^2)\cosh(\kappa_{zf}\lambda)
    -i2\kappa_{zf}k_z\sinh(\kappa_{zf}\lambda)]}{[(\kappa_{zf}+ik_z)^2e^{-\kappa_{zf}\lambda}-
    (\kappa_{zf}-ik_z)^2e^{\kappa_{zf}\lambda}]^2}.
\end{align}
\end{widetext}
Then the fermionic one-body tunneling strength becomes 
\begin{align}
    \mathcal{T}_{f,\bm{k}}\simeq T_{f,\bm{k}}\epsilon_k+V_f\mathcal{C}_{f,k_z}.
\end{align}
Simliar we can obtain the goldstino tunneling coupling strength as
\begin{align}
    \mathcal{G}_{Q,\bm{k}}=U_{bf}\operatorname{Re}\big[T_{b,\bm{k}}T^*_{f,\bm{k}}\big].
\end{align}
According to Eq.~(\ref{Tk}), we have
\begin{widetext}
\begin{align}
    \operatorname{Re}[T_{\alpha,\bm{k}}]=\frac{4k_z^2\kappa_{z\alpha}^2
    \cos(k_z\lambda)\cosh(\kappa_{z\alpha}\lambda)-2k_z\kappa_{z\alpha}
    (\kappa_{z\alpha}^2-k_z^2)\sin(k_z\lambda)\sinh(\kappa_{z\alpha}\lambda)}
    {(\kappa_{z\alpha}^2-k_z^2)^2\sinh^2(\kappa_{z\alpha}\lambda)+4\kappa_{z\alpha}^2
    k_z^2\cosh^2(\kappa_{z\alpha}\lambda)},
\end{align}
\begin{align}
    \operatorname{Im}[T_{\alpha,\bm{k}}]=-\frac{2k_z\kappa_{z\alpha}
    (\kappa_{z\alpha}^2-k_z^2)\cos(k_z\lambda)\sinh(\kappa_{z\alpha}\lambda)+
    4k_z^2\kappa_{z\alpha}^2\sin(k_z\lambda)\cosh(\kappa_{z\alpha}\lambda)}
    {(\kappa_{z\alpha}^2-k_z^2)^2\sinh^2(\kappa_{z\alpha}\lambda)+4\kappa_{z\alpha}^2
    k_z^2\cosh^2(\kappa_{z\alpha}\lambda)}.
\end{align}
\end{widetext}
Then the Goldstino tunneling strength can be calculated as 
\begin{align}
    \frac{\mathcal{G}_Q}{U_{bf}}=\operatorname{Re}[T_{b,\bm{k}}]\operatorname{Re}
    [T_{f,\bm{k}}]+\operatorname{Im}[T_{b,\bm{k}}]\operatorname{Im}[T_{f,\bm{k}}].
\end{align}

\section{Goldstino spectrum}\label{appendixB}

In this appendix, we show the calculation details of the Goldstino propagator and the feature of its spectrum. Generally, the explicit form of a one-loop Goldstino propagator is given by
\begin{align}\label{Lindhard}
    \Pi_{\bm{p}}(\Omega)=-\int\frac{d^3\bm{k}}{(2\pi)^3}
    \frac{f_f(\xi_{\bm{k}+\bm{p},f})+f_b(\xi_{\bm{k},b})}{\Omega+i\delta-\xi_{\bm{k}+\bm{p},f}+\xi_{\bm{k},b}},
\end{align}
where $\xi_{\bm{k},f}=\bm{k}^2/(2m_f)-\mu_f+\Sigma_{f}$ and $\xi_{\bm{k},b}=\bm{k}^2/(2m_b)-\mu_b+\Sigma_{b}$. Due to the different self-energy for condensate bosons and excited bosons, Eq.~(\ref{Lindhard}) can be rewritten as $\Pi_{\bm{p}}=\Pi_{\bm{p}}^p+\Pi_{\bm{p}}^c$, where the expressions of $\Pi_{\bm{p}}^p$ and $\Pi_{\bm{p}}^c$ are given by Eq.~(\ref{Pip}) and Eq.~(\ref{Pic}). When the temperature approaches zero, $N^0_b\rightarrow N_b$, while the distribution of bosons with $k>0$ vanishes, and 
\begin{align}\label{Pic0}
    \Pi^{\rm c}_{\bm{p}}(\Omega)\rightarrow -\int\frac{d^3\bm{k}}{(2\pi)^3}
    \frac{{f}_f(\xi_{\bm{k}+\bm{p},f})}{\Omega+i\delta-\xi_{\bm{k}+\bm{p},f}+\xi_{\bm{k},b}}.
\end{align}
It can be rewritten as 
\begin{align}
     \Pi^{\rm c}_{\bm{p}}(\Omega)=-\int\frac{d^3\bm{k}}{(2\pi)^3}\bigg[&\frac{{f}_f(\xi_{\bm{k},f})}{\Omega+i\delta-\xi_{\bm{k},f}+\xi_{\bm{k}-\bm{p},b}}\cr
     +&\frac{{f}_b(\xi_{\bm{k},b})}{\Omega+i\delta-\xi_{\bm{k}+\bm{p},f}+\xi_{\bm{k},b}}\bigg].
\end{align}
we define $\tilde{k}=k/k_{\rm F}$, $\tilde{\Omega}=\Omega/\epsilon_{\rm F}$, and $\tilde{\mu}_{Q}=\mu_{Q}/\epsilon_{\rm F}$, with $\epsilon_{\rm F}$ and $k_{\rm F}$ denoting the Fermi energy and Fermi momentum of fermions. After conducting the angular integral, we have 
\begin{align}
    \Pi^{\rm c}_{\bm{p}}(\Omega)&=\frac{k_{\rm F}^3}{8\pi^2\epsilon_{\rm F}\tilde{p}}\int d\tilde{k}\tilde{k}\cr
    \bigg[&{f}_f(\xi_{\bm{k},f})\ln\bigg(\frac{\tilde{\Omega}+i\delta-2\tilde{k}\tilde{p}+\tilde{p}^2+\tilde{\mu}'_{Q}}{\tilde{\Omega}+i\delta+2\tilde{k}\tilde{p}+\tilde{p}^2+\tilde{\mu}'_{Q}}\bigg)\nonumber\\
    +&{f}_b(\xi_{\bm{k},b})\ln\bigg(\frac{\tilde{\Omega}+i\delta-2\tilde{k}\tilde{p}-\tilde{p}^2+\tilde{\mu}'_{Q}}{\tilde{\Omega}+i\delta+2\tilde{k}\tilde{p}-\tilde{p}^2+\tilde{\mu}'_{Q}}\bigg)\bigg],
\end{align}
where $\mu'_{Q}=\mu_{Q}-(\Sigma_f-\Sigma_b)=\epsilon_{\rm F}+U_{bb}N_b$. Due to the infinitesimally small number $\delta$, the formulas inside the parentheses can be rewritten as 
\begin{align}
    &\frac{\tilde{\Omega}+i\delta-2\tilde{k}\tilde{p}\pm\tilde{p}^2+\tilde{\mu}'_{Q}}{\tilde{\Omega}+i\delta+2\tilde{k}\tilde{p}\pm\tilde{p}^2+\tilde{\mu}'_{Q}}\cr
    =&~\frac{\big|(\tilde{\Omega}\pm\tilde{p}^2+\tilde{\mu}'_{Q})^2-4\tilde{k}^2\tilde{p}^2\big|}{(\tilde{\Omega}+2\tilde{k}\tilde{p}\pm\tilde{p}^2+\tilde{\mu}'_{Q})^2}\cr
    & \times \exp\bigg\{i\arctan\bigg[\frac{4\tilde{k}\tilde{p}\delta}{(\tilde{\Omega}\pm\tilde{p}^2+\tilde{\mu}'_{Q})^2-4\tilde{k}^2\tilde{p}^2}\bigg]\bigg\}\cr
    =&~A^\pm_{\tilde{\bm{p}}}(\tilde{\Omega},\tilde{k})e^{i\tilde{\theta}^\pm_{\bm{p}}(\tilde{\Omega},\tilde{k})}.
\end{align}
Then the real part of the Goldstino propagator reads
\begin{align}
    \operatorname{Re}\Pi^{\rm c}_{\bm{p}}(\Omega)=\frac{3N_f}{4\epsilon_{\rm F}\tilde{p}}\int &d\tilde{k}\,\tilde{k}\big\{{f}_f(\xi_{\bm{k},f})\ln\big[A^+_{\tilde{\bm{p}}}(\tilde{\Omega},\tilde{k})\big]\cr
    &+{f}_b(\xi_{\bm{k},b})\ln\big[A^-_{\tilde{\bm{p}}}(\tilde{\Omega},\tilde{k})\big]\big\},
\end{align}
where the fermionic density reads $N_f=k_{\rm F}^3/6\pi^2$.

For the imaginary part of $\Pi$, according to the identity $\frac{1}{\Gamma+i\delta}=\mathcal{P}\frac{1}{\Gamma}-i\pi\delta(\Gamma)$, we have $\operatorname{Im}\Pi_{\bm{p}}=\operatorname{Im}\Pi^{\rm p}_{\bm{p}}+\operatorname{Im}\Pi^{\rm c}_{\bm{p}}$, where 
\begin{align}\label{ImPip}
    \operatorname{Im}\Pi^{\rm p}_{\bm{p}}(\Omega)=\pi N^0_b\delta(\Omega+\epsilon_{\rm F}-\bm{p}^2/2m)
\end{align}
and
\begin{align}\label{ImPic}
    \operatorname{Im}\Pi^{\rm c}_{\bm{p}}(\Omega)&=\pi\int\frac{d^3\bm{k}}{(2\pi)^3}\big[{f}_f(\xi_{\bm{k}+\bm{p},f})+{f}_b(\xi_{\bm{k},b})\big]\cr
    &\delta\big[\Omega-(\bm{k}+\bm{p})^2/2m+\bm{k}^2/2m+\mu'_{Q}\big].
\end{align}
By setting the direction of $\bm{p}$ along the $z$-axis, taking $\int\frac{d^3\bm{k}}{(2\pi)^3}\rightarrow\int dk d\theta\, k^2\sin\theta$, and replacing $\cos\theta$ by a parameter $q=|\bm{k+\bm{p}}|$, the equation above can be reformulated as
\begin{align}
    \operatorname{Im}\Pi^{\rm c}_{\bm{p}}(\Omega)&=\frac{k_{\rm F}^3}{4\pi\epsilon_{\rm F}\tilde{p}}\int\tilde{k}d\tilde{k}\int_{|\tilde{k}-\tilde{p}|}^{\tilde{k}+\tilde{p}}\tilde{q}d\tilde{q}\nonumber\\
    \times\big[{f}_f&(\xi_{\bm{q},f})+{f}_b(\xi_{\bm{k},b})\big]\delta\big(\tilde{\Omega}-\tilde{q}^2+\tilde{k}^2+\tilde{\mu}'_{Q}\big),
\end{align}
where $\tilde{k}$ and $\tilde{\Omega}$ are normalized by $k_{\rm F}$ and $\epsilon_{\rm F}$. By using the property of the delta function $\delta[f(x)]=\delta(x-x_0)/|f'(x_0)|$, with $f(x_0)=0$, we have
\begin{align}
    \operatorname{Im}\Pi^{\rm c}_{\bm{p}}(\Omega)=&~\frac{k_{\rm F}^3}{8\pi\epsilon_{\rm F}\tilde{p}}\int\tilde{k}d\tilde{k}\int_{|\tilde{k}-\tilde{p}|}^{\tilde{k}+\tilde{p}}d\tilde{q}\nonumber\\
    &\times\big[{f}_f(\xi_{\bm{q},f})+{f}_b(\xi_{\bm{k},b})\big]\delta(\tilde{q}-\tilde{q}_0),
\end{align}
where $\tilde{q}_0=\sqrt{\tilde{\Omega}+\tilde{k}^2+\tilde{\mu}'_{Q}}$. Meanwhile, for the integral to be nonzero, $\tilde{q}_0$ must satisfy the inequality $|\tilde{k}-\tilde{p}|\leq \tilde{q}_0\leq \tilde{k}+\tilde{p}$, leading to a lower limit for the integral over $k$ as
\begin{align}
    \tilde{k}\geq\frac{1}{2}\bigg|\frac{\tilde{\Omega}+\tilde{\mu}'_{Q}}{\tilde{p}}-\tilde{p}\bigg|.
\end{align}
Therefore, we obtain the expression for the imaginary part of the Lindhard function as given by
\begin{subequations}
    \begin{align}
    \operatorname{Im}\Pi^{\rm p}_{\bm{p}}(\Omega)=\pi N^0_b\delta(\Omega+\epsilon_{\rm F}-\bm{p}^2/2m),
\end{align}
\begin{align}
    \operatorname{Im}\Pi^{\rm c}_{\bm{p}}(\Omega)=\frac{3\pi N_f}{4\epsilon_{\rm F}\tilde{p}}\int_{\alpha}^\infty\tilde{k}d\tilde{k}\big[{f}_f(\xi_{q_0,f})+{f}_b(\xi_{k,b})\big],
\end{align}
\end{subequations}
where $\alpha=\frac{1}{2}\big|\frac{\tilde{\Omega}+\tilde{\mu}'_{Q}}{\tilde{p}}-\tilde{p}\big|$.

By applying the random phase approximation (RPA), the diagrams with the same order as the one-loop diagram should be summed, yielding the explicit Goldstino propagator:
\begin{align}
    \chi_{\bm{p}}(\Omega)=\frac{\Pi_{\bm{p}}(\Omega)}{1+U_{bf}\Pi_{\bm{p}}(\Omega)}.
\end{align}
The Goldstino spectral function is given by the imaginary part of the propagator:
\begin{align}
    \operatorname{Im}\tilde{\chi}_{\bm{p}}(\Omega)=\frac{\operatorname{Im}
    \tilde{\Pi}_{\bm{p}}(\Omega)}{[1+\tilde{U}_{bf}\operatorname{Re}\tilde{\Pi}^{\rm c}_{\bm{p}}(\Omega)]^2+[\tilde{U}_{bf}\operatorname{Im}\tilde{\Pi}^{\rm c}_{\bm{p}}(\Omega)]^2},
\end{align}
where $\tilde{U}_{ij}=2U_{ij}N_f/\epsilon_{\rm F}=8a_{ij}k_{\rm F}/(3\pi)$, $\tilde{\chi}_{\bm{p}}=\chi_{\bm{p}}\epsilon_{\rm F}/(2N_f)$, and $\tilde{\Pi}_{\bm{p},i}=\Pi_{\bm{p}}\epsilon_{\rm F}/(2N_f)$. 
\begin{figure*}[t]
    \centering
    \includegraphics[width=15cm]{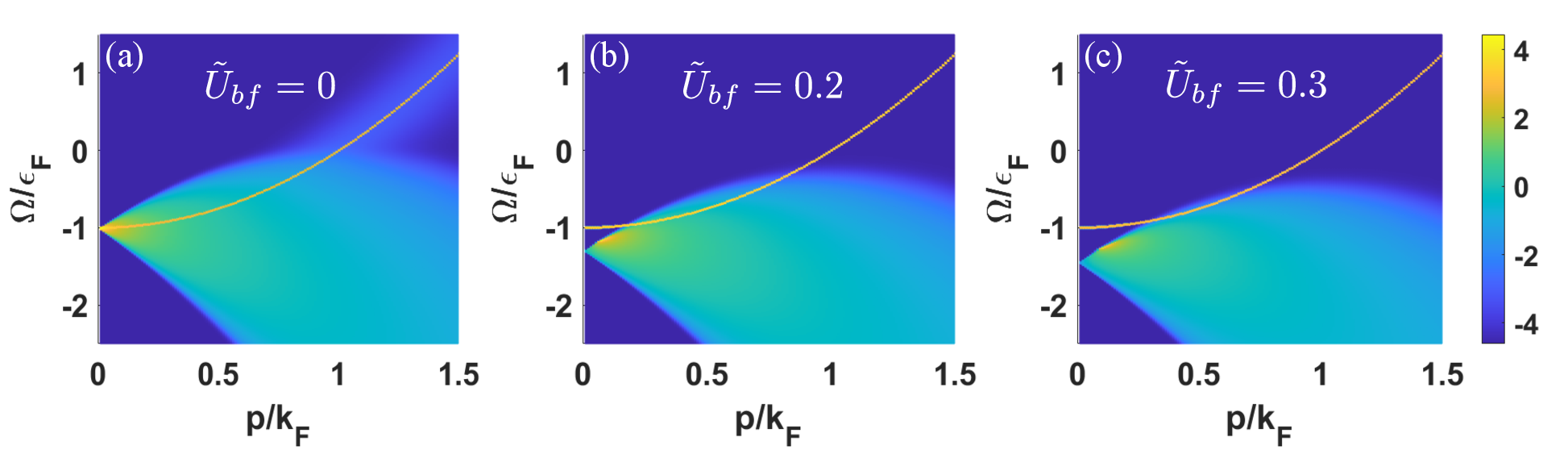}
    \caption{Goldstino spectra with different interaction strengths in a Bose-Fermi mixture with explicitly broken supersymmetry. The color bars show the logarithmic scale. We fix the ratio $N_b/N_f=2$ and the interaction parameters are chosen to be the experimental values of the $^{173}$Yb-$^{174}$Yb mixture, $a_{bf}/a_{bb}\simeq1.32$. The Goldstino poles (bright yellow line) originates from the BEC phase, and the continuum (light area) originates from the excitation of bosons. 
    }\label{imchi}
\end{figure*}

Fig.~\ref{imchi} shows the Goldstino spectra with different interaction strengths in a $^{173}$Yb-$^{174}$Yb mixture. The poles denoted by a sharp peak arise from the condensate of bosons and yield a nonzero energy gap, which is caused by the supersymmetry breaking associated with the chemical potential bias between fermions and bosons. The continuum comes from the exchange of fermions and free bosons. The sharp-peaked structure indicates that it is a long-life collective excitation.

\section{Modification from Bogoliubov modes}\label{appendixC}

In this Appendix, we estimate the additional modification for one-loop Goldstino propagator given by the low-momentum Bogoliubov sound modes. The Bogoliubov transformation reads
\begin{align}
    &b_{\bm{k}}=u_{\bm{k}}\alpha_{\bm{k}}+v^*_{-\bm{k}}\alpha^\dagger_{-\bm{k}},\cr
    &b^\dagger_{\bm{k}}=u^*_{\bm{k}}\alpha^\dagger_{\bm{k}}+v_{-\bm{k}}\alpha_{-\bm{k}},
\end{align}
where $\alpha_{\bm{k}}$ and $\alpha^\dagger_{\bm{k}}$ are annihilation and creation operators of Bogoliubov modes. The coefficients are given by 
\begin{align}
    u^2_{\bm{k}}=&\frac{1}{2}\bigg[\frac{(\bm{k}^2/(2m)+U_{bb}N_b)}{E_{\bm{k}}}+1\bigg],\cr
    v^2_{\bm{k}}=&\frac{1}{2}\bigg[\frac{(\bm{k}^2/(2m)+U_{bb}N_b)}{E_{\bm{k}}}-1\bigg],
\end{align}
where the dispersion of the Bogoliubov mode reads 
\begin{align}
    E_{\bm{k}}=\sqrt{\frac{\bm{k}^2}{2m}\bigg(\frac{\bm{k}^2}{2m}+2U_{bb}N_b\bigg)}.
\end{align}
Notice that for small momentum, this dispersion relation becomes linear: $E_{\bm{k}}\simeq c|\bm{k}|$, where $c\equiv \sqrt{U_{bb}N_b/m}$. At large momentum, the interaction effect becomes negligible and the dispersion remains that of a free boson: $E_{\bm{k}}\simeq\bm{k}^2/(2m)$. Thus we consider the affect of Bogoliubov excitations at low momentum: $|\bm{k}|<k_c$, with $k_c\equiv\sqrt{2mU_{bb}N_b}$ denoting the characteristic momentum at which the behavior of the spectrum changes from linear to quadratic.

The modification term for Goldstino one-loop propagator by the Bogoliubov modes is given by 
\begin{align}
    \Pi^\alpha_{\bm{p}}(\Omega)=-\int\frac{d^3\bm{k}}{(2\pi)^3}\bigg[&u_{\bm{k}-\bm{p}}^2\frac{f_f(\xi_{\bm{k},f})}{\Omega-\xi_{\bm{k},f}+E_{\bm{k}-\bm{p}}}\cr
    +&v_{\bm{k}-\bm{p}}^2\frac{1-f_f(\xi_{\bm{k},f})}{\Omega-\xi_{\bm{k},f}-E_{\bm{k}-\bm{p} }}\bigg],
\end{align}
where $\xi_{\bm{k},f}=\bm{k}^2/(2m_f)-\mu_f+U_{bf}N_b$. In the weak-coupling regime, the characteristic momentum is much smaller than the Fermi momentum: $k_c\ll k_{\rm F}$, making the numerator of the second term an extremely small value at small momentum ($|\bm{k}|<k_c$). Therefore the contribution of the second term is negligible in the low-momentum region. Then we consider, for instance, the case $|\bm{p}|=0$, where the contribution of the soft modes ($|\bm{k}|<k_c$) to $\Pi^\alpha_{\bm{p}}$ reduces to
\begin{align}
    \Pi^\alpha_{\bm{0}}(\Omega)=-\frac{1}{2\pi^2}\int_0^{k_c}dk\,k^2\frac{u_{\bm{k}}^2}{\Omega-\xi_{\bm{k},f}+E_{\bm{k}}}.
\end{align}
We note that in the integration region, $u^2_{\bm{k}}$ is of order $UN/(c|\bm{k}|)$ and the denominator of the integrand is of order $UN$ when $\Omega$ is small. Combining these estimates, we eventually find that $\Pi^\alpha$ is of order $k_c^2/c\approx m^{3/2}\sqrt{UN}$. On the other hand, according to Eq.~(13), when $f_b(\xi_{\bm{k},b})\rightarrow 0$ we have
\begin{align}
    \Pi^{\rm c}_{\bm{0}}(\Omega)=-\frac{1}{2\pi^2}
    \frac{N_f}{\Omega-\xi_{\bm{k},f}+\xi_{\bm{k},b}},
\end{align}
which is of order $1/U$. Therefore as long as $U_{bb}$ and $U_{bf}$ are small, the modification by the Bogoliubov mode is much smaller than the contribution from the momentum region $|\bm{k}|>k_c$ with mean-field correction, as considered in the main text.

\bibliography{ref.bib}

\end{document}